\documentclass[%
aip,
reprint,
nofootinbib,
amsmath,amssymb,
floatfix
]{revtex4-2}

\usepackage[pdftex]{graphicx}
\usepackage{dcolumn}
\usepackage{bm}

\usepackage[T2A]{fontenc}  
\usepackage[utf8]{inputenc} 
\usepackage[english]{babel} 

\usepackage[colorlinks=true,linkcolor=blue]{hyperref}

\usepackage{mathptmx}
\usepackage{etoolbox}
\usepackage{enumitem}

\newcommand{\un}[1]{\>\mathrm{#1}}

\DeclareMathOperator{\erf}{erf}

\makeatletter
\def\@email#1#2{%
 \endgroup
 \patchcmd{\titleblock@produce}
  {\frontmatter@RRAPformat}
  {\frontmatter@RRAPformat{\produce@RRAP{*#1\href{mailto:#2}{#2}}}\frontmatter@RRAPformat}
  {}{}
}%
\makeatother
\begin{document}



\title{Subsonic and supersonic gas flows to condensation surface}

\author{A.~P.~Kryukov}
\affiliation{National Research University ``Moscow Power Engineering Institute'', 111250 Moscow, Russia}

\author{V.~V.~Zhakhovsky}
\email{basi1z@ya.ru}
\affiliation{Joint Institute for High Temperatures of Russian Academy of Sciences, 125412 Moscow, Russia}

\author{V.~Yu.~Levashov}
\affiliation{Institute of Mechanics, Lomonosov Moscow State University, 119192 Moscow, Russia}%

\date{\today}

\begin{abstract}
Intense heat-mass transfer in a gas flow to a condensation surface is studied with the consistent atomistic and kinetic theory methods.
The simple moment method is utilized for solving the Boltzmann kinetic equation (BKE) for the nonequilibrium gas flow and its condensation, while molecular dynamics (MD) simulation of a similar flow is used for verification of BKE results.
We demonstrate that BKE can provide the steady flow profiles close to those obtained from MD simulations in both subsonic and supersonic regimes of steady gas flows. Surprisingly, the elementary theory of condensation is shown with BKE results to have a good accuracy in a wide range of gas flow parameters.

MD confirms that a steady supersonic gas flow condensates on a surface at the distinctive temperature after formation of a standing shock front in reference to this surface, which can be interpreted as a permeable condensating piston.
The last produces the shock compression but completely absorbs incoming gas flow in contrast to a common impermeable piston. The shock front divides the vapor flow on the supersonic and subsonic zones, and condensation of compressed gas happens in the subsonic regime. The complete and partial condensation regimes are discussed. It is shown that above the certain surface temperatures determined by the shock Hugoniot the runaway shock front stops an inflow gas and condensation is ceased.
\end{abstract}

\maketitle

\section{Introduction}

Evaporation and condensation processes underlie many critical technologies, in which the various heat and mass transfer apparatus, air separation plants, refrigerators and heat pump condensers are used. These processes plus sublimation and desublimation are involved in chemical vapor deposition --- this method is used for formation of films and deposited micro-structures with specified properties.
Evaporation of droplets of different sizes and condensation on them both in pure vapor and in vapor-gas mixtures also attracts considerable interest in propulsion engineering for a long time~\cite{Lamanna:2020}. The basic processes involved in thermophysics of liquid-vapor phenomena, including vapor condensation of both low-melting and high-temperature metals, separation of gas mixtures and removal of harmful impurities from environment by condensation followed by sublimation are reviewed in~\cite{Carey:2020}.
Systematic analysis of condensation may be thought of as beginning from the Nusselt's pioneer work \cite{Nusselt:1916} published a century ago. Thus, the mass and heat transfer at condensation is studied for a long time, but until now the development of new technologies require profound understanding the associated processes evolving in new environments.

\begin{figure}[t]
\centering\includegraphics[width=0.95\columnwidth]{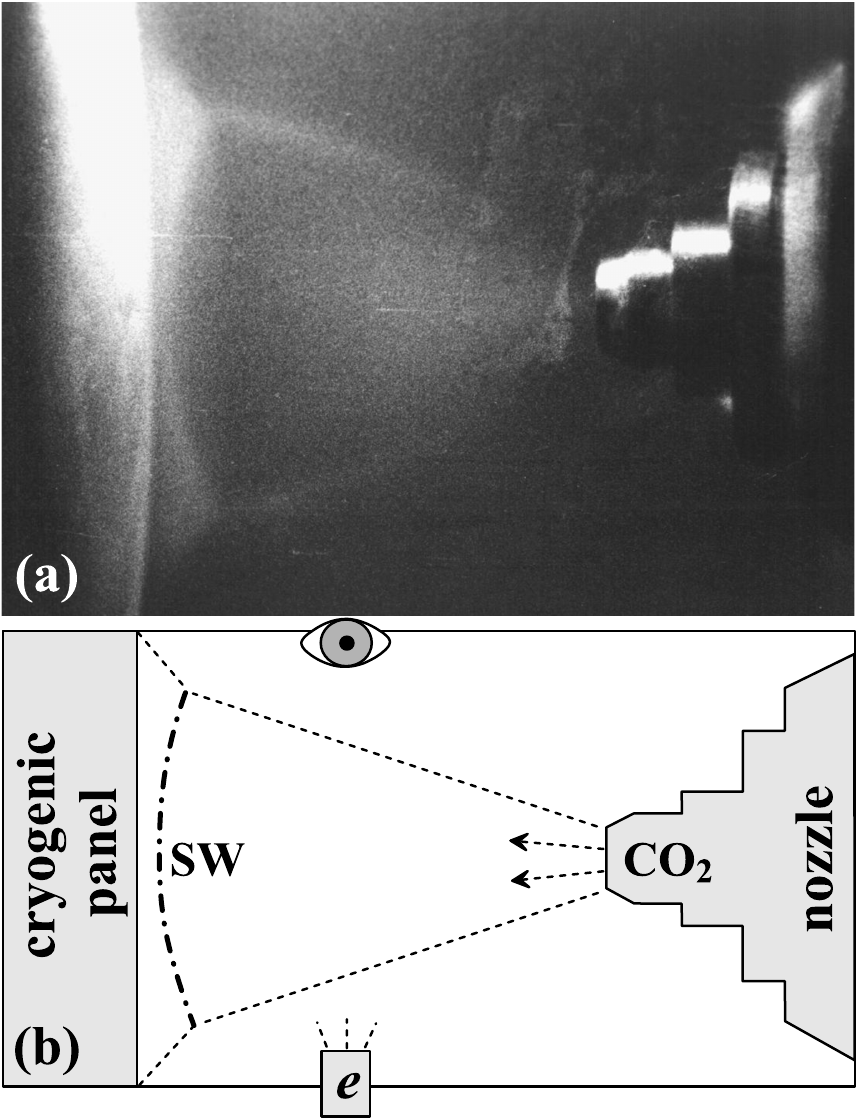}
\caption{\label{fig:exp} Condensation of $\un{CO_2}$ jet from nozzle on cryogenic panel in vacuum chamber --- \textbf{(a)} snapshot and \textbf{(b)} scheme of experiment. $\un{CO_2}$ gas is compressed by a quasi-stationary shock wave (SW) near the panel. The plasma electron gun (e) is used to visualize gas distribution by light emission from the excited molecules. Gas is imaged through a window located at the angle of $90^{\circ}$ to the electron gun.}
\end{figure}

In the sixties of the last century the development of cryo-vacuum equipment for imitations of gas flows around a spacecraft from the continuum to free-molecular regime was called for the new space technologies.
The demand for vacuum pumping has aroused considerable interest in closer studies of strong non-equilibrium condensation processes.
These processes are characterized by high rate of heat and mass transfer. Figure~\ref{fig:exp}(b) illustrates the scheme of our experiment performed in 1989 for observing condensation of carbon dioxide jet on the cryo-panel surface. The corresponding photo (previously unpublished) obtained by shooting through a vacuum chamber window is presented in Fig.~\ref{fig:exp}(a). The vacuum system maintains the pressure near 1 Pa in the experimental chamber during experiment, while the pressure about $1.2\times 10^5\un{Pa}$ is supported in a stagnation chamber.

The supersonic gas flow is formed as a result of outflow through a Laval nozzle from the stagnation chamber. The Laval nozzle throat has diameter of $0.98\un{mm}$ and the half-opening angle of $12^\circ$. The distance from the nozzle to a cryogenic panel, where condensation of incoming gas proceeds, is 60 mm, see geometry of the experi of incoming flow mental chamber in Fig.~ \ref{fig:exp}. The plasma electron gun is used to visualize density distribution in the gas flow. The electron collisions with molecules result in their  ionization or excitation followed by light emission due to optical transitions to a ground energy level. The emitted light has higher intensity from regions with higher local gas density.

The initial surface temperature of cryogenic panel before opening of the Laval nozzle is equal $79-80\un{K}$, which is low enough to cause desublimation of $\un{CO_2}$ molecules from incoming gas flow. It is observed that a layer of increased density is appeared in front of the panel at 5 second after experiment starting, and it is observed throughout the entire time until the gas supply is stopped after 20 seconds. The gas distribution imaged just before this time is presented in Fig.~\ref{fig:exp}(a).
Thickness of the high-density layer outlined by SW in Fig.~\ref{fig:exp}(b) increases slowly with time. Thus, it may be concluded from these observations that a quasi-stationary shock wave outgoing from the condensation surface with speed slightly above the $\un{CO_2}$ flow velocity is formed, and it continues until the flow is cut off.

Since application of the common quasi-equilibrium continual approaches is unreasonable in highly nonequilibrium conditions the methods of molecular-kinetic theory were invoked to study the high-rate condensation processes.
Application of the Boltzmann kinetic equation (BKE) and its simplified models in such conditions gave new insight into the processes associated with evaporation and condensation.
As a result, the quantitative descriptions of many experiments were obtained, and the ranges of various regimes of evaporation/condensation processes were determined \cite{Kryukov:2021}.

However, the BKE methods require the complicated boundary conditions including the velocity distribution functions of molecules moving from the interface surface. Early in the history of BKE application, such information was provided by empirical or semiempirical methods --- by setting the unknown evaporation and condensation coefficients, using an experimental saturation line of gas evaporated from a condensed phase, choosing a simple interaction model of vapor molecules with the condensation surface, such as a diffuse or mirror reflection rules.
The development of molecular dynamics (MD) method makes it possible to avoid such simplifying assumptions and empiricism, and perform direct atomistic simulations of evaporation and condensation processes, as well as develop an unified end-to-end technique for study the corresponding processes in a two-phase system consisting of a condensed phase, its vapor, and an interface layer between them. The active works are carrying out in this direction during recent years, see Refs.~\cite{Frezzotti:2011,Frezzotti:2019,Busuioc:2020,Kon:2014,Kobayashi:2017}.

Recent progress in cluster computer technology and a rapid rise of computational capability together with development of high-performance parallel MD codes have led to widespread use of MD method for simulation of complex physical processes in fields, which were difficult to imagine two decades ago. It awakes our interest in direct MD simulation of high-rate condensation, which were actively investigated by methods of molecular-kinetic theory 30--40 years ago. Along with rigorous BKE and MD methods the approximate analytical approaches are of interest, because they allow to estimate the various characteristics of evaporation and condensation processes without any computation costs.

In this paper, a comprehensive approach is proposed to obtain stationary solutions in both subsonic and supersonic regimes of condensation on a flat surface using the relatively simple moment method for BKE. Large-scale MD simulations of the same problems are performed to validate the results obtained by the moment method. In particular, our MD results confirm that the supersonic condensation with a shock wave standing ahead the surface is correctly described by the moment method. Also we demonstrate that the elementary theory of condensation works well in a wide range of gas flow parameters.

\section{Numerical solution of the Boltzmann kinetic equation and molecular dynamics}
\label{sec:methods}

The problem setup illustrated by Fig.~\ref{fig:exp}(b) is the same for both numerical techniques -- BKE and MD methods. The liquid (or solid) layer or surface is located on the left side of simulation domain. The temperature of this surface $T_s$ is fixed or should be determined, while the density corresponding to this temperature on a saturation line is known, see Fig.~\ref{fig:Ns}. Gas with the given atom density $n_{in}$ and temperature $T_{in}$ inflows with the velocity $u_{in}$ in a simulation domain from the right. It is required to find a stationary solution of such condensation problem, which provides the spatial profiles (including density, temperature, and mass velocity profiles) of a steady flow within the simulation domain. It is also required to demonstrate that the obtained solution remains the same if the length of simulation domain increases. The details of both methods and consistency conditions between them are discussed below.

\subsection{MD simulation of steady condensation}

MD simulations of two-phase systems consisting of gas and condensed phase of argon-like atoms interacting via a smoothed Lennard-Jones (L-J) potential \cite{Zhakhovsky:1999} were performed in the same way that described in \cite{Zhakhovsky:2018}.
The width of computational domain $L_x=200\un{nm}$ was chosen to be much larger than a mean free path in the gas phase, so that the Knudsen layer thickness at the left condensation surface could not reach the right boundary. Periodic conditions were imposed along the transverse axes, along which the MD box dimensions $L_y=L_z=100\un{nm}$ were chosen large enough to increase the atom statistics to make smoother the $x$-axis profiles. The number of atoms is not fixed in simulations, but on an average the MD box contains the order of $~2\times 10^6$ atoms. MD simulations were performed with our in-house parallel code MD-$\mathrm{VD^3}$ using the Voronoi dynamic domain decomposition \cite{Zhakhovsky:2005,Egorova:2019}.

Two Langevin thermostats are used to establish the steady gas flow, respectively, to feed atoms at the right boundary and remove atoms from the simulation domain at the left boundary.
The application zones of these thermostats are indicated by gray areas in some figures presented in this paper. From the right boundary, a layer of gas with the given density $n_{in}$ and mass flow velocity $u_{in}$ is inserted into the right thermostat zone of 20 nm thick, in which thermalization of gas atoms to the given temperature $T_{in}$ and flow velocity took place.
Another thermostat of 5 nm thick at the left boundary cools the incoming atoms to $T_{out}$ and decelerates their averaged velocity to the required output speed $u_{out}$ in order to create a condensed phase layer in which the atoms slowly drift to the left boundary, where they are removed from the simulation domain.

The output speed $u_{out}$ is controlled by a total number of atoms in the computational domain. The feedback algorithm adjusts $u_{out}$ so as to set the desired atom number. A steady flow and condensation are established after a certain number of iterations of the algorithm, with the thickness of condensed phase becoming larger than the left thermostat zone and stabilized. A transition interface layer, where atomic processes of condensation and evaporation actually occur, is established between the condensed phase and  flowing gas. After reaching the stationary regime, the accumulation of statistics begins and the time-averaged profiles of physical quantities are constructed along $x$-axis.

\begin{figure}[t]
\centering\includegraphics[width=1.\columnwidth]{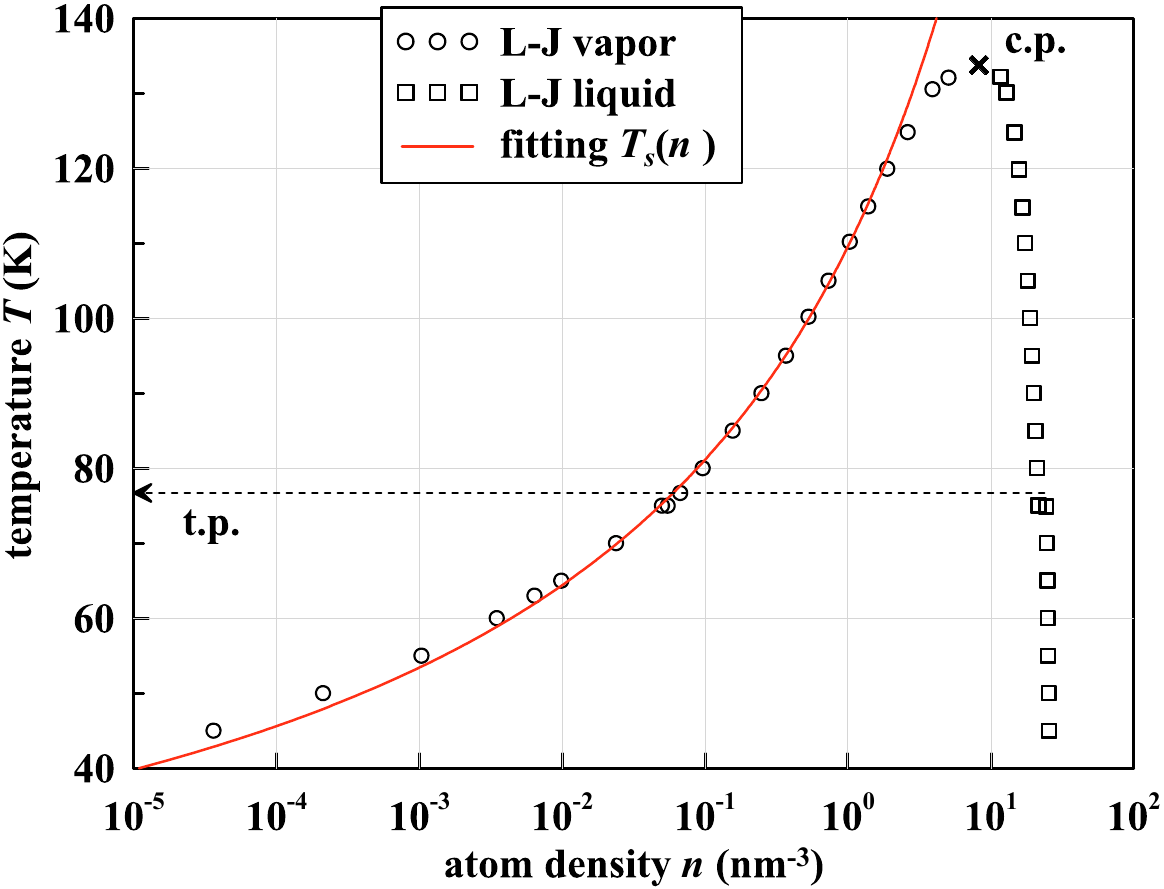}
\caption{\label{fig:Ns}
Liquid-vapor equilibrium line of an argon-like system obtained from MD simulations using the smoothed $\un{L-J}$ potential. The condensed phase is represented by squares and the gas phase by circles. The triple point $T_\mathrm{tp}=76.7\un{K}$ and an estimate of the critical point (c.p.) are shown. The saturated vapor temperature $T_s(n)$ is fitted by the red curve using Eq.~(\ref{eq:Ns}).}
\end{figure}

To make a direct comparison of two solutions of gas flow condensation problem obtained by the BKE and MD methods, it is necessary for BKE to set a boundary condition on the condensation surface using the saturation line $n_s(T)$ calculated from MD simulations.
Obtaining a vapor phase in equilibrium with a condensed phase was performed as described in Ref.~\cite{Zhakhovsky:2018}.
The atom density of saturated vapor shown in Fig.~\ref{fig:Ns} is fitted by a function
\begin{equation}
\label{eq:Ns}
n_s = \exp \left(a-\frac{E_v}{k_B T}\right)= \exp \left(a-\frac{T_v}{T} \right),
\end{equation}
where the heat of vaporization is written in the form $E_v=k_B T_v$ to get rid of the Boltzmann constant further, $a=6.57$ and $T_v=720\un{K}$ are the fitting parameters.
Figure~\ref{fig:Ns} demonstrates a good accuracy of this fitting formula almost up to the critical temperature.

\subsection{BKE and moment method for steady condensation}

The Boltzmann kinetic equation (BKE) in the absence of mass forces can be written in a general form as follows
\begin{equation}
\label{eq:Boltz}
\frac{\partial f}{\partial t}+\mathbf{\xi}\frac{\partial f}{\partial \mathbf{r}}=J,
\end{equation}
where $f\left( \mathbf{r},t,\mathbf{\xi}\right)$ is a velocity distribution function, $\mathbf{r}(x,y,z)$ are the Cartesian coordinates, $t$ is time, $\mathbf{\xi}\left(\xi_x,\xi_y,\xi_z\right)$ is the molecule velocity in the laboratory coordinate system, $J=\int\limits^\infty_0 \int\limits^{2\pi}_0 \int\int\limits^{+\infty}_{-\infty}\int\left(f^\prime f^\prime_1-f f_1\right)|\overrightarrow{g}| b db d\varepsilon d\overrightarrow{\xi_1}$ is the collision integral. Other notations in the Boltzmann equation are traditional as in~\cite{Kogan:1969}.
One-dimensional stationary problem is considered, in which Eq.~(\ref{eq:Boltz}) takes the form
$\xi_x (\partial f/\partial x)=J.$

The six-moment method for solving the BKE utilizes the two-stream Maxwell approximation $f=f_1+f_2$ of the velocity distribution function:
\begin{equation}
\label{eq:f1f2}
\begin{cases}
f_1=\frac{n_1(x)}{(2 \pi R T_1(x))^{3/2}} \exp\left[\frac{\left(\xi_x-u_1(x)\right)^2+\xi_y^2+\xi_z^2}{2 R T_1(x)}\right],\;\; \xi_x>0, \\
f_2=\frac{n_2(x)}{(2 \pi R T_2(x))^{3/2}} \exp\left[\frac{\left(\xi_x-u_2(x)\right)^2+\xi_y^2+\xi_z^2}{2 R T_2(x)}\right],\;\; \xi_x<0,
\end{cases}
\end{equation}
where the right components $n_1,T_1,u_1$ (for $\xi_x>0$)  and the left components $n_2,T_2,u_2$  (for $\xi_x<0$), corresponding to atom density, temperature and mass flow velocity, are functions of $x$ only.

To obtain a system of moment equations the both sides of BKE must be multiplied by several distinctive functions $\varphi$, and then be integrated in a velocity space. The following functions $\varphi_i$ are chosen:
\begin{displaymath}
\varphi_1=1, \; \varphi_2=\xi_x, \; \varphi_3=\mathbf{\xi}^2, \;  \varphi_4=\xi_x^2, \; \varphi_5=\xi_x^3, \; \varphi_6=\xi_x \mathbf{\xi}^2
\end{displaymath}
Thus, six equations for six unknown moment functions $M_i$ are obtained instead one BKE.
\begin{equation}
\label{eq:moment}
\frac{d M_i\left(n_1,T_1,u_1,n_2,T_2,u_2\right)}{dx}=I_i\left(n_1,T_1,u_1,n_2,T_2,u_2\right),
\end{equation}
where  $M_i=\int \varphi_i f d\mathbf{\xi}=\int\limits^\infty_{-\infty}\int\limits^\infty_{-\infty}\int\limits^\infty_{-\infty} \varphi_i f d\xi_x d\xi_y d\xi_z$  are $i\in[1,6]$ moments of the distribution function, $I_i=\int \varphi_i J d\mathbf{\xi}=\int\limits^\infty_{-\infty}\int\limits^\infty_{-\infty}\int\limits^\infty_{-\infty} \varphi_i J d\xi_x d\xi_y d\xi_z$  are moments of the collision integral. The latter are calculated for Maxwell molecules, because only for such an interaction potential these integrals can be calculated analytically. This facilitates greatly calculation of the collision integral moments, since it reduces to finding out the moments of distribution function.

The left boundary condition at $x=0$ must determine a velocity distribution function $f_1$ for atoms moving from the condensation surface after evaporation or reflection, while $f_2$ for incoming atoms should be obtained via solution of the moment equations (\ref{eq:moment}). Assuming the diffuse nature of reflection and evaporation processes it can be written as follows
\begin{equation}
\label{eq:bound1}
   f_1|_{x=0}=\frac{n_s}{(2 \pi R T_s)^{3/2}} \exp\left[\frac{\xi_x^2+\xi_y^2+\xi_z^2}{2 R T_1(x)}\right],\;\; \xi_x>0,
\end{equation}
where $u_1(0)=0$, and $n_1(0)=n_s,$ $T_1(0)=T_s$, if the evaporation and condensation coefficients are assumed to equal unity. Here $n_s$ is the atom density of saturated vapor at the given surface temperature $T_s$. As this takes place, the foregoing distribution function determines a counterflow of gas evaporated at the given temperature $T_s$, and thus it defines the left boundary condition at the condensation surface.
In this work the saturation line $n_s(T)$ of argon-like two-phase system was fitted by Eq.~(\ref{eq:Ns}) to MD simulation results, see Fig.~\ref{fig:Ns}.

For the right boundary condition at $x \rightarrow \infty$ it is supposed that:
\begin{equation}
\label{eq:bound2}
   f|_{x\rightarrow \infty}=\frac{n_\infty}{(2 \pi R T_\infty)^{3/2}} \exp\left[\frac{\left(\xi_x^2-u_\infty\right)^2+\xi_y^2+\xi_z^2}{2 R T_\infty(x)}\right],
\end{equation}
where $n_1=n_2=n_\infty,$ $T_1=T_2=T_\infty,$ $u_1=u_2=u_\infty$ are used. So the functions $f_1$ and $f_2$ has the same form as in Eq.~(\ref{eq:bound2}), but their domains of definition are separated by $\xi_x=0$ according to Eq.~(\ref{eq:f1f2}).

\begin{figure}[t]
\centering\includegraphics[width=1.\columnwidth]{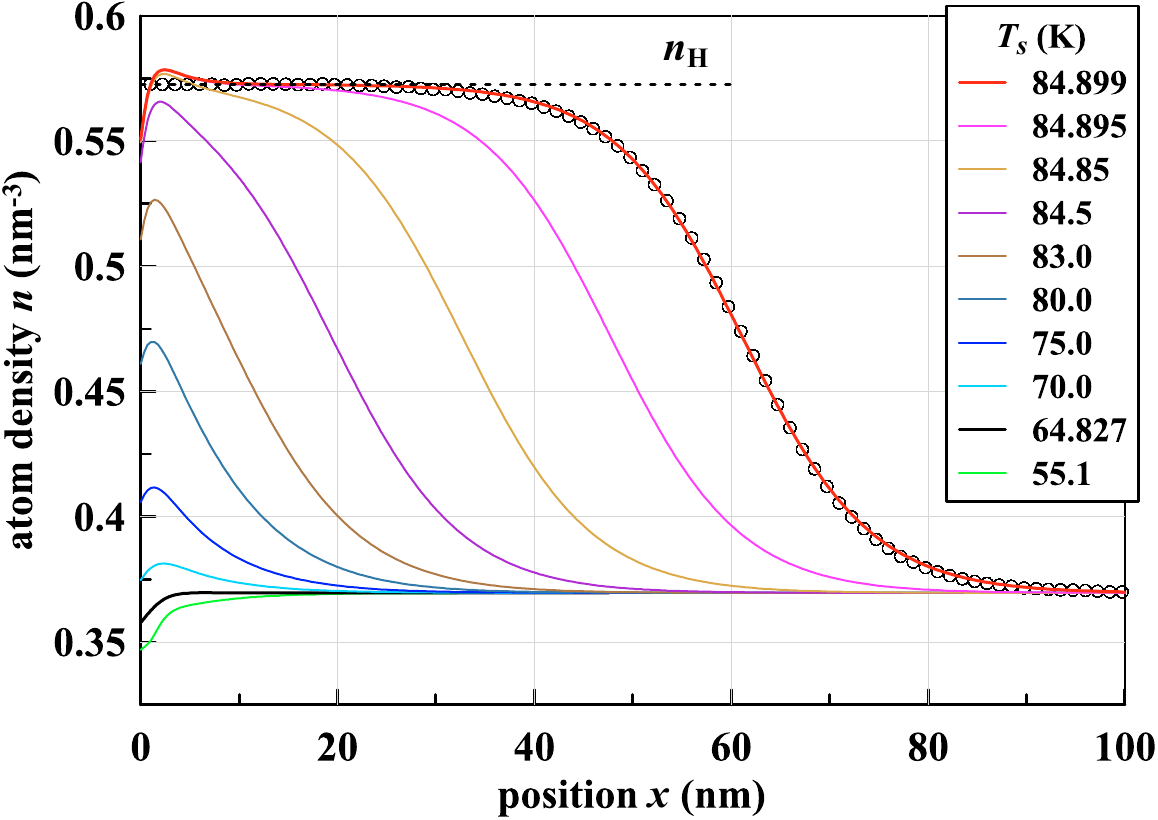}
\caption{\label{fig:Ncomp_ss} Atom density profiles calculated by BKE method for steady condensation of gas flow with supersonic velocity $u_\infty=-250\un{m/s}$ on cryogenic surface $x=0$ at different temperatures. Steady flows exist only for $T_s \leq 84.899\un{K}$, and a steady SW profile is approached at the upper limit of $T_s$. MD simulation of L-J gas provides a SW profile (shown by circles) scaled to the true density $n_{\mathrm{H}}=0.57262\un{nm^{-3}}$ of shock-compressed perfect gas (a dashed plateau). }
\end{figure}

The dimensionless variables are introduced as follows: $n^{\star}=n/n_s,$ $T^{\star}= T/T_s,$ $u^{\star}= u/\sqrt{2 R T_s},$ $\xi_x^{\star}= \xi_x/\sqrt{2 R T_s}$.
Hereinafter the stars are omitted, and the moment equations Eqs.~(\ref{eq:a1}--\ref{eq:a6}) are derived in Appendix. Such a system of equations with the boundary conditions Eqs.~(\ref{eq:bound1},\ref{eq:bound2}) was formulated and presented originally by reports \cite{Hatakeyama:1979,Oguchi:1980}, but in Appendix this system is written in another form.
Comparison of early solutions of strong condensation problems for subsonic and supersonic flows obtained by the moment method \cite{Kryukov:1985,Kryukov:1991}, and with a model kinetic equation~\cite{Bishaev:1973,Sone:1986,Aoki:1989}, and using the direct simulation Monte Carlo (DSMC) \cite{Abramov:1989,Abramov:1990} demonstrates that those solutions agree closely with each other.

Numerical integration of the boundary value problem for the system of ordinary differential equations (ODE) Eqs.~(\ref{eq:moment}) was performed using the left boundary condition as an initial condition for the Cauchy problem.
The ODE are integrated by means of fourth order Runge-Kutta formulae with adjustable step to control the accuracy \cite{Press:1992}. Shooting method was realized by minimization of a target function, composed of deviations of integration results from the right boundary condition, with the use of the downhill simplex algorithm \cite{Nelder:1965,Press:1992} combined with random walks in multi-dimensional space of unknown initial values at the left boundary.

Since the boundary placed at infinity is inaccessible for numerical integration of Eqs.~(\ref{eq:moment}) the calculations are performed on a sequence of the bounded segments along the $x$-axis.
After finding a solution on a given segment, the right boundary condition is moved further away from the left boundary, and a new solution on a larger segment must be found. This procedure is repeated until the difference between the successive solutions becomes sufficiently small.

For comparison of BKE solutions with MD simulations a transport cross-section for the Maxwell's molecules is adjusted to fit MD simulation of steady SW profile. Here the thickness of shockwave front is used for spatial scaling the BKE profiles with MD profiles. The mean free path $l=2.3791\un{nm}$ is found to provide a good fit with MD simulation of SW propagating with speed $u_s=250\un{m/s}$ in the L-J gas having the initial atom density $n_\infty=0.3696\un{nm^{-3}}$ and temperature $T_\infty=95\un{K}$, as shown in Fig.~\ref{fig:Ncomp_ss}.
Such an initial gas state is chosen since it is used for direct comparisons of MD and BKE results. The corresponding transport cross-section $\sigma=1.1373\un{nm^2}$ for Maxwell's molecules at the given $T=95\un{K}$ is used in our BKE code to provide the mean free path $l=1/(n \sigma)$ for arbitrary initial densities.

\section{Condensation of supersonic gas flow}
\label{sec:supersonic}

Stationary solutions of the moment Eqs.~(\ref{eq:a1}--\ref{eq:a6}) for a supersonic flow with velocity $u_\infty=-250\un{m/s}$ ($M_\infty=1 4$), $T_\infty=95\un{K}$ and $n_\infty=0.3696\un{nm^{-3}}$ incoming to the condensation surface at different $T_s$ are presented in Fig.~\ref{fig:Ncomp_ss}. Among the many solutions, there is one at $T_s = 64.827\un{K}$ for which the incoming flow has minimal perturbations in the surface vicinity. The presented density profiles also show that the condensing flow accelerates, rarefies, and cools as it approaches the surface at $T_s< 64.827\un{K}$, while at larger $T_s$ it decelerates and compresses.
The density and temperature of gas increases, and the occupied region expands until the gas state approaches a shock-compressed state at the shock Hugoniot, in other words in a shock wave staying at rest in the incoming stream (i.e. moving with velocity $u_s=|u_\infty|$ in a static uncompressed gas. Such a state is feasible as $T_s=84.899\un{K}$ is approached, above which there are no stationary solutions of the momentum equations. It means physically that at larger $T_s$ a SW running from the surface with a speed greater than the flow velocity is formed, which proves the impossibility of forming a stationary flow at such surface temperatures.

\begin{figure}[t]
\centering\includegraphics[width=1.\columnwidth]{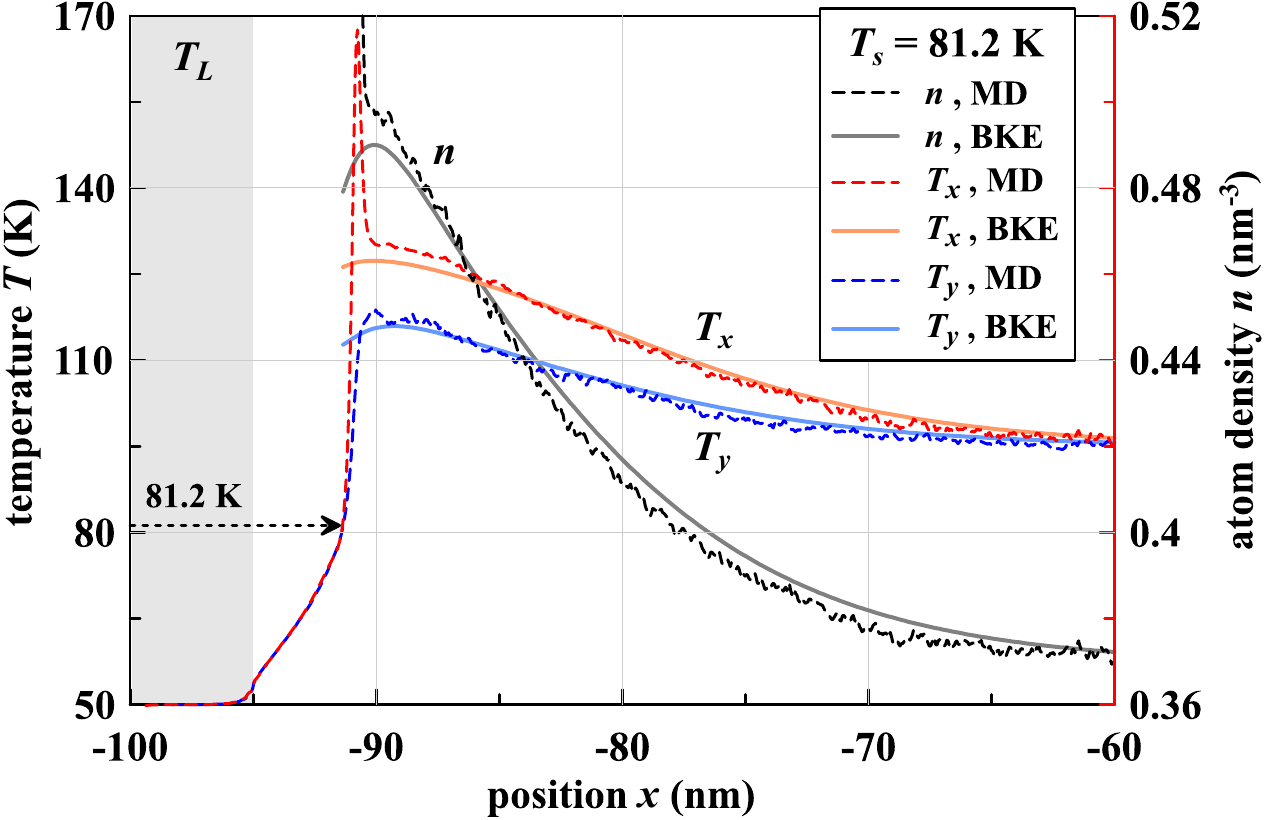}
\caption{\label{fig:nT_81K} Atom density, longitudinal and transverse temperatures profiles for steady condensation with supersonic velocity $u_{in}=-250\un{m/s}$ on condensation surface at $T_s=81.2\un{K}$ obtained by MD (dashed lines) and BKE (solid lines) methods. The left thermostat with $T_L=50\un{K}$ is applied for atoms in the gray zone. The cold surface of BKE profiles is shifted to $T_s$  position within an inter-phase layer in MD profile. }
\end{figure}

The stationary solution with a SW front obtained by the momentum method gives a flow profile on the spatial coordinate normalized by the mean free path. For direct comparison with MD results, we fitted the path length so that the SW profiles from the MD and BKE methods coincide, as seen in Fig.~\ref{fig:Ncomp_ss}.
Before fitting the shock-compressed density of L-J gas $n_{\mathrm{H}}=0.60286\un{nm^{-3}}$ is scaled down to $n_{\mathrm{H}}=0.57262\un{nm^{-3}}$ obtained from the shock Hugoniot for the perfect gas with $\gamma=5/3$.

It should be noted that the stationary solutions obtained by the moment method can only approximate the true state of shock-compressed gas, but never reaching it. In other words, the density profile at $T_s=84.899\un{K}$ shown in Fig.~\ref{fig:Ncomp_ss} corresponds to a not true shock wave, since the apparent coincidence of the compressed gas density with the exact density $n_{\mathrm{H}}=0. 57262\un{nm^{-3}}$ (and no temperature $T_{\mathrm{H}}=130.04\un{K}$ shown), calculated by the well-known Rankine-Hugoniot jump conditions for the perfect monatomic gas, is deceptive. In reality, the presented density and temperature profiles have a small negative slope ($dn/dx < 0$ and $dT/dx < 0$) even at the apparent plateau, otherwise the flow parameters would stop changing after reaching a true plateau with zero derivatives in the moment method. To find the exact $T_s$ corresponding to a true shock-compressed gas condensation, one should not perform an all-in-one calculation of supersonic flow, but first calculate the subsonic flow velocity and the shock-compressed gas parameters behind the SW front having the velocity of this flow, and then apply the moment method to calculate the subsonic flow condensation. In this approach, the shock front stays at rest ahead the condensation surface at arbitrary but not small distance determined by the transient processes of establishing the stationary flow. The results obtained by this approach are given in \ref{sec:linear}.

The shock front cuts the flow in two zones --- a supersonic flow ahead the front and a subsonic flow behind, where the last may reach the condensation surface. There is a single stationary solution of the momentum equations for the zone with a subsonic shock-compressed gas flow. Finding a unique solution of these equations is discussed in the next section \ref{sec:subsonic}. Therefore, for a fixed SW velocity $u_s>|u_\infty|$ moving away from the surface, one can again obtain a stationary profile of the condensing gas issuing from the shock front, if the last runs away from the surface far enough to avoid interference with flow variations caused by condensation. It is easy to see that as the SW speed increases, the gas flowing through the shock front will increasingly decelerate (in the reference frame in Fig.~\ref{fig:Ncomp_ss}) and the compressed gas stops completely at some $u_s$. Such a trivial solution providing a non-condensing motionless shock-compressed gas can be obtained only at $P_s=P^*_{\mathrm{H}}$, where $P_s$ is the saturated vapor pressure at the surface temperature $T_s$, and $P^*_{\mathrm{H}}$ is the pressure in the shock-compressed gas staying at rest with respect to the surface. For such a case, the evaporating surface acts as a piston, which generates vapor with the necessary pressure $P^*_{\mathrm{H}}$ to balance with the shock-compressed gas.

\begin{figure}[t]
\centering\includegraphics[width=1.\columnwidth]{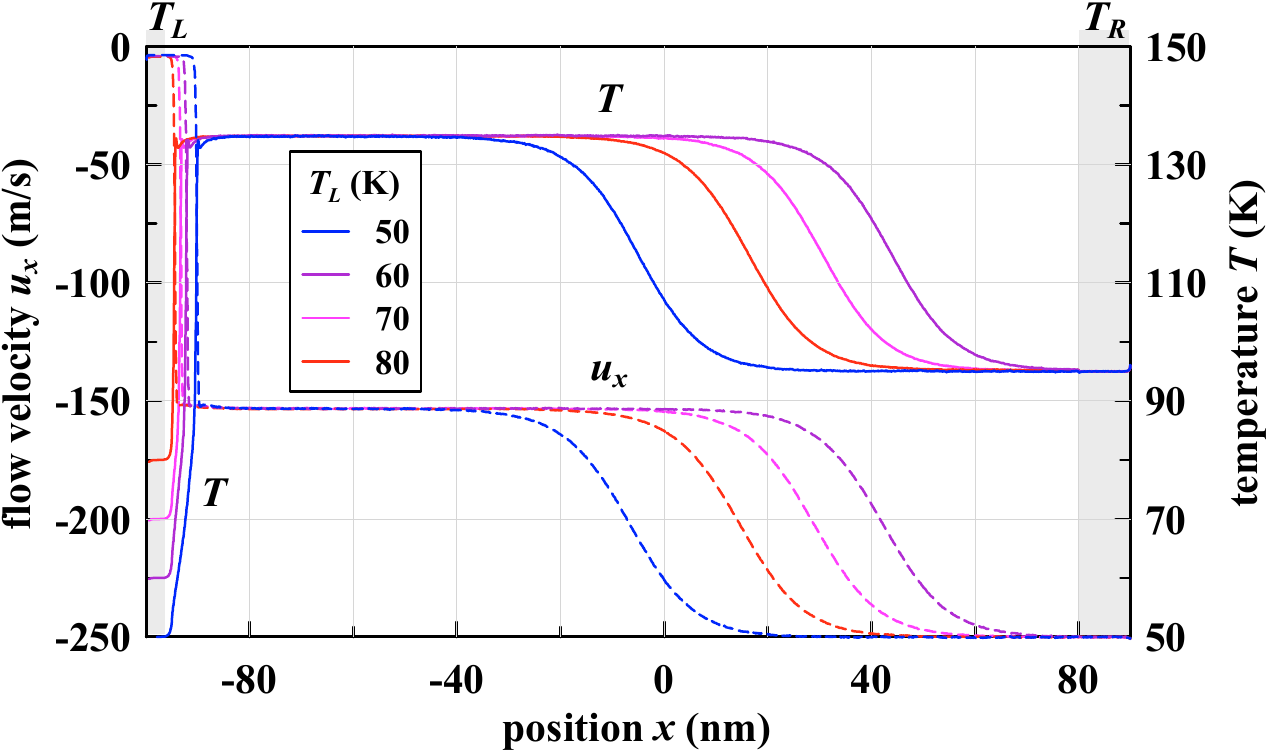}
\caption{\label{fig:Ux-T_sw} Temperature and flow velocity profiles between the condensation surface (left) and the input thermostat with $T_R=95\un{K}$ in MD simulation of supersonic flow with $u_{in}=-250\un{m/s}$. Parameters of shocked gas are independent of the left thermostat $T_L$. Positions of standing shock fronts are determined only by a total number of atoms sustained in simulation. }
\end{figure}

Using the well-known shock Hugoniot of perfect gas with $\gamma=5/3$ it is easy to obtain all characteristics of the subsonic gas flow behind the shock front, if the required deceleration of this flow is given. For a gas incoming to the shock front with the parameters shown in Fig.~ ~\ref{fig:Ncomp_ss}, the initial flow velocity must slow down by $|u_\infty|=250\un{m/s}$ to completely stop a flow behind the front.
This condition is satisfied by a SW with $u_s=413.1\un{m/s}$ (relative to a motionless uncompressed gas), $n_{\mathrm{H}}=0.9361\un{nm^{-3}}$ and $P^*_{\mathrm{H}}=3016\un{kPa}$. The evaporating surface of argon-like liquid, which has the condensation curve shown in Fig.~\ref{fig:Ns}, must be at $T^*_s=120.5\un{K}$ to evaporate the perfect gas with the same pressure.
Thus, the boundary of complete cessation of  condensation is determined by the shock Hugoniot of condensing gas. This boundary as a function of flow velocity is given in \ref{sec:linear}.

\begin{figure}[t]
\centering\includegraphics[width=1.\columnwidth]{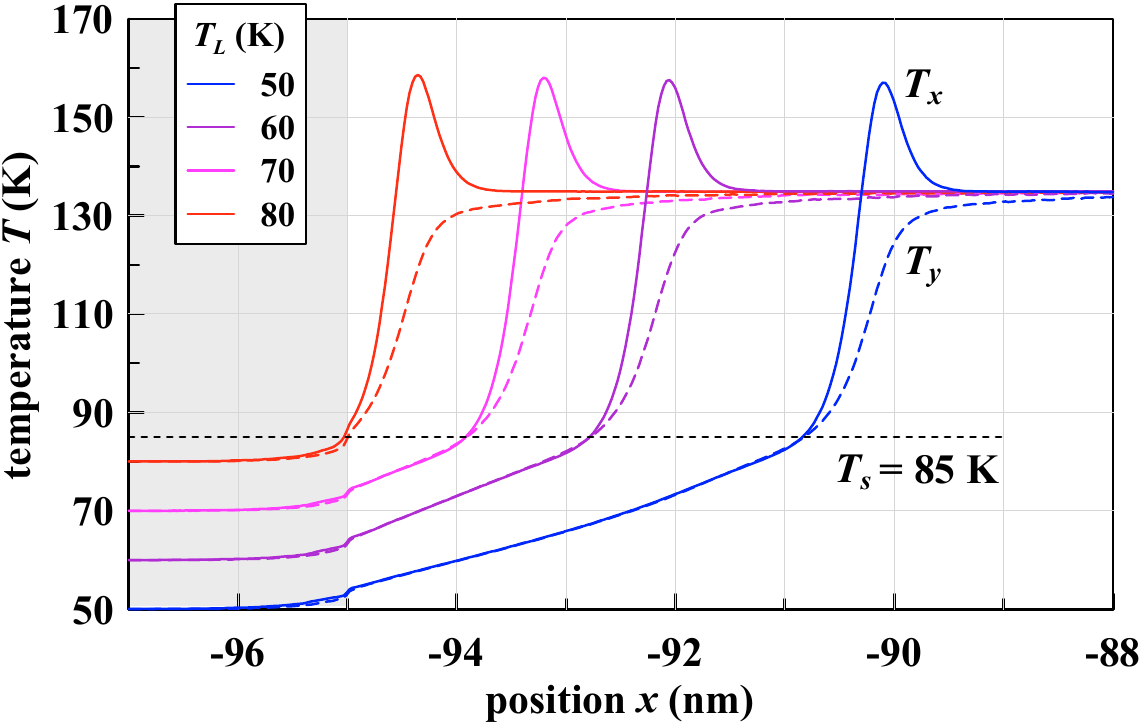}
\caption{\label{fig:Ux-T_sw-cond} Longitudinal $T_x$ and transverse $T_y$ temperatures near condensation surface obtained from MD simulation of supersonic flow with $u_{in}=-250\un{m/s}$. All different steady SW profiles are established at the same $T_s=85\un{K}$ independently of the left thermostat $T_L$, see also Fig.~\ref{fig:Ux-T_sw}.}
\end{figure}

For MD simulation of gas condensation, the cryogenic panel was represented not as a perfect surface in BKE method, but as a layer of condensed matter, the leftmost part of which is maintained at a given temperature $T_L<T_s$ using the left Langevin thermostat, whose area is colored gray in Fig.~\ref{fig:nT_81K}.
During condensation, the heat of vaporization is released, and the temperature increases with distance from the thermostat area. Therefore, as the thickness of condensed phase increases, so does the temperature in the transition surface layer, where we define the splitting point of the $T_x$ and $T_y$ profiles as the surface temperature $T_s$ \cite{Zhakhovsky:2018}. The thickness of condensed phase depends on the output velocity $u_{out}$ in the left Langevin thermostat, which controls the total number of atoms in the computational domain. The feedback algorithm adjusts $u_{out}$ so as to establish a steady regime of condensation. In such a regime, the thickness of condensed phase stabilizes, which allows $T_s$ to reach some stationary value too.

To generate a gas flow with the given mass velocity $u_{in}$ and temperature $T_{in}=T_\infty$
the right Langevin thermostat is used, in which a thin layer of gas with the corresponding parameters is inserted at the right boundary. The range of application of this thermostat is indicated by the right gray area in Fig.~\ref{fig:Ux-T_sw}.
If the simulation domain is long enough the flow variations initiated by condensation at its left boundary cannot reach its right boundary because the acoustic perturbations are carried downstream (to the left) by the supersonic flow.

The comparison of the flow profiles obtained at $T_s=81.2\un{K}$ and supersonic $u_\infty=u_{in}=-250\un{m/s}$ by the BKE and MD calculations are shown in Fig.~\ref{fig:nT_81K}.
The condensation surface (left boundary) for the BKE profiles is shifted to the position of condensed phase surface determined in MD simulation. A good agreement of the longitudinal and transverse temperature profiles, as well as density, can be seen in the Knudsen layer ahead of the condensation surface. A strong difference arises only in the transient interface layer, which appears naturally in MD simulations but is nonexistent in the BKE method.

\begin{figure}[t]
\centering
\includegraphics[width=1.\columnwidth]{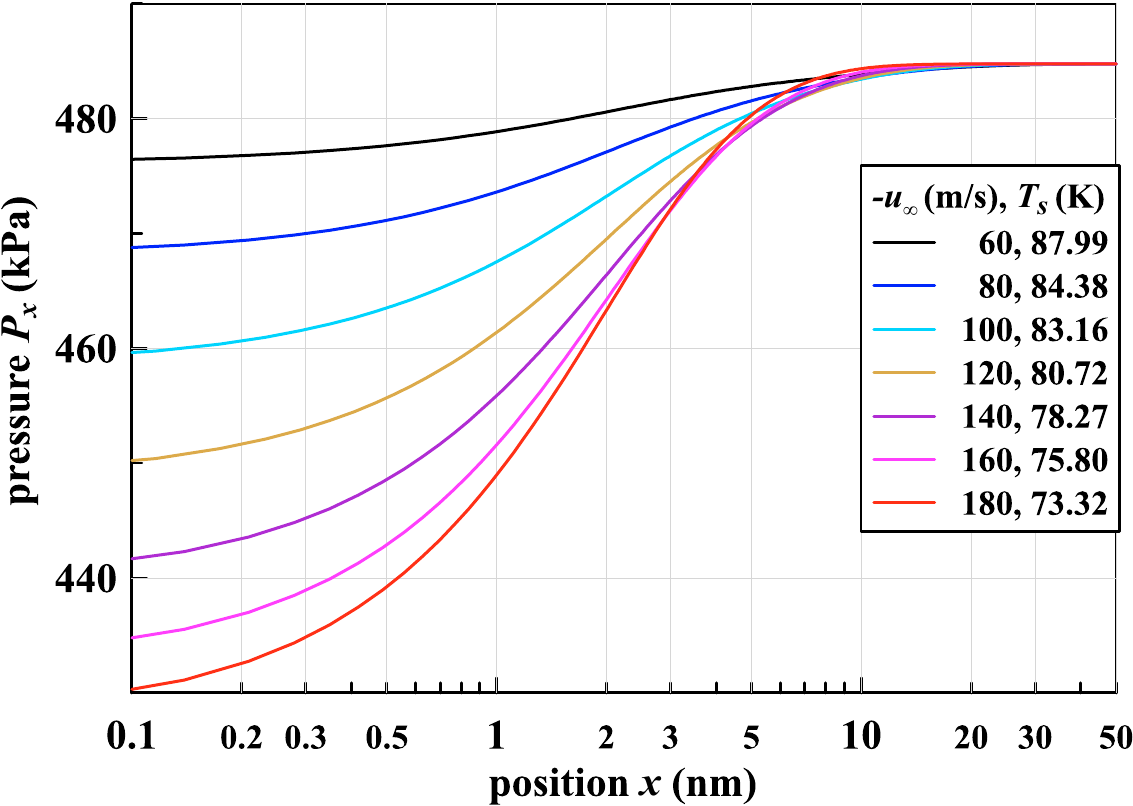}
\caption{\label{fig:P_subson} Steady profiles of longitudinal pressure $P_x=n k_B T_x$ obtained for gas flow with $T_\infty=95 \un{K}$, $n_\infty=0.3696\un{nm^{-3}}$ and given subsonic velocities $u_\infty$ towards the condensation surface $x=0$ at different temperatures $T_s$.}
\end{figure}

The results of MD simulation of supersonic condensation with formation of steady shock waves in the inflowing supersonic gas with $M_\infty=1.4$ and $T_{in}=95\un{K}$, and for different temperatures $T_L$ of condensed phase (solid at 50 K, and liquid at 60,70,80 K) are presented in Figs.~\ref{fig:Ux-T_sw} and \ref{fig:Ux-T_sw-cond}. These results confirm that the steady shock waves can be formed in a gas flow upstream of the condensation surface, as it is predicted by the molecular kinetic theory.

A steady flow regime with a standing shock wave obtained at different $T_L$ using the feedback algorithm described above can lead to different positions of the shock front, as can be seen from Fig.~\ref{fig:Ux-T_sw}. Those positions are determined only by a target number of atoms that the algorithm tries to maintain in the domain, and therefore the established position of steady SW front is arbitrary. The target number is chosen so that the front position will be in the central part of computational domain. The SW speed and the shock-compressed gas parameters do not depend on the algorithm for obtaining a stationary profile, and are identical in all four cases. These parameters are determined by the Rankine-Hugoniot conditions for a shock wave propagating with velocity $u_s=|u_{in}|$ in a given L-J gas.

The steady condensation is independent of the left thermostat temperature $T_L$ applied to atoms within a gray zone of the fixed thickness shown in Fig.~\ref{fig:Ux-T_sw-cond}, because an additional condensed phase layer with increasing temperature accumulates over the thermostat zone. The incoming gas condenses on this additional layer until its thickness reaches a value required to establish the same surface temperature $T_s$ independent of $T_L$. Figure~\ref{fig:Ux-T_sw-cond} demonstrates the growth of temperature in such condensed phase layers of different thicknesses beyond the left thermostat, and the constancy of $T_s=85\un{K}$.

The condensation process develops in the transition interface layer between gas and liquid, which thickness is about 1 nm, but this layer is replaced by a surface with zero thickness in the BKE calculations. A large jump of nonequilibrium $T_x$ occurs in this layer,  the origin of which is discussed in \cite{Zhakhovsky:2018}.
It should also be noted that the equation of state of the L-J gas, as well as its shock Hugoniot, differs significantly from the corresponding properties of the perfect gas. For these reasons, we do not provide a direct comparison of the slightly different SW profiles obtained from the BKE and MD simulations, although the $T_s=84.9\un{K}$ required for steady condensation of the perfect gas turned out to be very close to $T_s=85\un{K}$ for condensation of the L-J gas with the same initial temperature, density and flow velocity (but with a slightly lower pressure).

\begin{figure}[t]
\centering\includegraphics[width=1.\columnwidth]{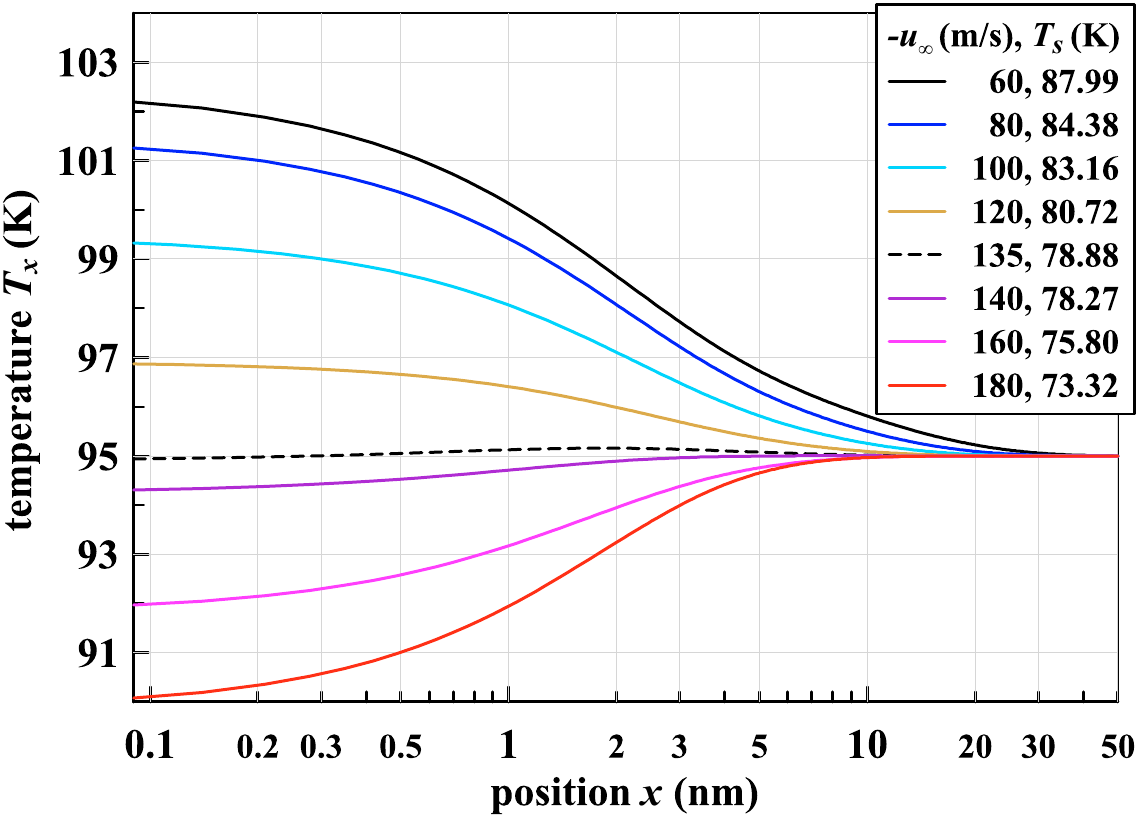}
\caption{\label{fig:Te_subson} Longitudinal temperatures $T_x$ near condensation surface with different temperatures $T_s$. Steady profiles were obtained by BKE for subsonic velocities $u_\infty$ of saturated gas flow at $T_\infty=95 \un{K}$. Black line shows the highest $T_x$ established at $|u_\infty|=60\un{m/s}$, after which $T_x$ near surface starts to decrease for lower flow velocities.}
\end{figure}

\section{Condensation of subsonic gas flow}
\label{sec:subsonic}

\begin{figure}[t]
\centering\includegraphics[width=1.\columnwidth]{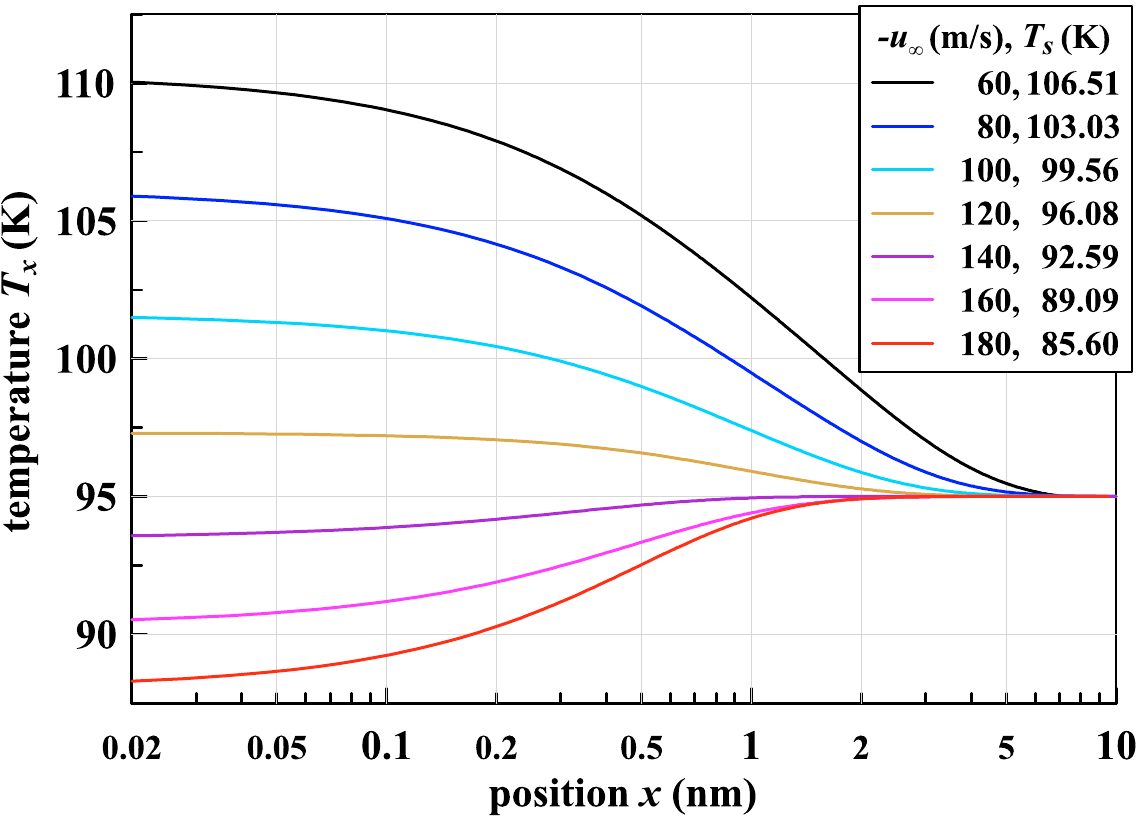}
\caption{\label{fig:Tx5_subson} Steady BKE profiles of longitudinal temperature $T_x$ near condensation surface with different temperatures $T_s$ for gas flows with subsonic velocities $u_\infty$. Gas has fivefold supersaturated vapor density $n_\infty=1.848\un{nm^{-3}}$ at $T_\infty=95\un{K}$. Condensation happens at hotter surfaces with $T_s > T_\infty$ for flow velocities $|u_\infty| < 126\un{m/s}$. }
\end{figure}

It is known \cite{Kryukov:1991,Labuntsov:1979,Kryukov:1985,Sone:1986,Aoki:1989,Abramov:1989,Abramov:1990} that the assignment of any two parameters of gas far away from a surface with the given temperature uniquely determines a third parameter required for steady condensation of one-dimensional subsonic flow, unlike for condensation of supersonic flow on such a surface.
For example, an unknown flow velocity can be uniquely determined for a steady flow of gas with the given temperature and pressure. In this work we apply the fitting method described in Section \ref{sec:methods} to find such a single numerical solution of the boundary value problem for the system of moment Eqs.~(\ref{eq:a1}--\ref{eq:a6}) given in Appendix. The obtained results are presented in Figs.~\ref{fig:P_subson}, \ref{fig:Te_subson}, and \ref{fig:Tx5_subson}. The reader should be warned that the pressure and temperature profiles shown here describe the average characteristics of nonequilibrium non-Maxwellian distribution functions for the longitudinal and transverse atom velocities, which depend on longitudinal spatial coordinates, but are time-independent due to stationarity of solutions. Therefore, the corresponding terms used below, such as longitudinal pressure, temperature, and sound velocity, should be perceived as formal extensions of their definitions to strongly nonequilibrium gas states.

In an incoming subsonic flow of the perfect gas with parameters $n_\infty=0. 3696\un{nm^{-3}}$ and $T_\infty=95 \un{K}$ the longitudinal pressure defined as $P_x=n k_B T_x$ monotonically decreases with approaching to the condensation surface. This gas state is chosen from the L-J vapor saturation curve in Fig.~\ref{fig:Ns} used to specify an evaporating counterflow from the condensation surface. For all initial gas velocities $u_\infty$ shown in Fig.~\ref{fig:P_subson}, there is a pressure drop, which causes a monotonic increase in flow velocity and a density drop as it approaches the condensation surface.
The thickness of the near-surface layer of gas in which the acceleration of the flow occurs is several tens of mean free paths of atoms --- in other words, it is the Knudsen layer.
Such acceleration can give a supersonic flow velocity if the initial flow velocity $u_\infty$ is sufficiently close to the speed of sound. For example, the flow velocity on a profile with $u_\infty=-180\un{m/s}$, shown in Fig.~\ref{fig:P_subson}, exceeds the local sound speed at $x=4.4\un{nm}$ and reaches $M_x=u_x/c_x=1.09$ at the condensation surface. Such a transition can theoretically lead to formation of a condensation shock in a supersonic flow of a real gas where it reaches a supersaturated state, but due to the small thickness of this zone, in which only a few interatomic collisions occur, formation of a condensation jump becomes impossible.

On the other hand, Fig.~\ref{fig:Te_subson} indicates that the longitudinal flow temperature $T_x$ can both increase as it approaches the condensation surface and decrease. The decrease of $T_x$ occurs at high velocities and low surface temperatures $T_s < 78.88\un{K}$. The increase in the longitudinal temperature is observed in the flow at relatively low velocities and large surface temperatures $T_s > 78.88\un{K}$.  This behavior is due to the competition of adiabatic cooling of the rarefying gas with the increasing influence of the evaporating countercurrent on the broadening of the non-Maxwell function of the atom longitudinal velocity distribution, and hence on the growth of $T_x$, near the condensation surface. The balance of these two factors gives a nearly isothermal flow profile with $T_x(x)\approx 95\un{K}$ at $T_s=78.88\un{K}$ and $u_\infty=-135\un{m/s}$. It is also interesting to note that the maximum $T_x\approx 102\un{K}$ is found on the profile with
$u_\infty=-60\un{m/s}$ and $T_s=87.99\un{K}$, shown in Fig.~\ref{fig:Te_subson}. At lower flow rates, the temperature $T_x$ begins to decrease and tends to $T_\infty=95 \un{K}$.

\begin{figure}[t]
\centering\includegraphics[width=1.\columnwidth]{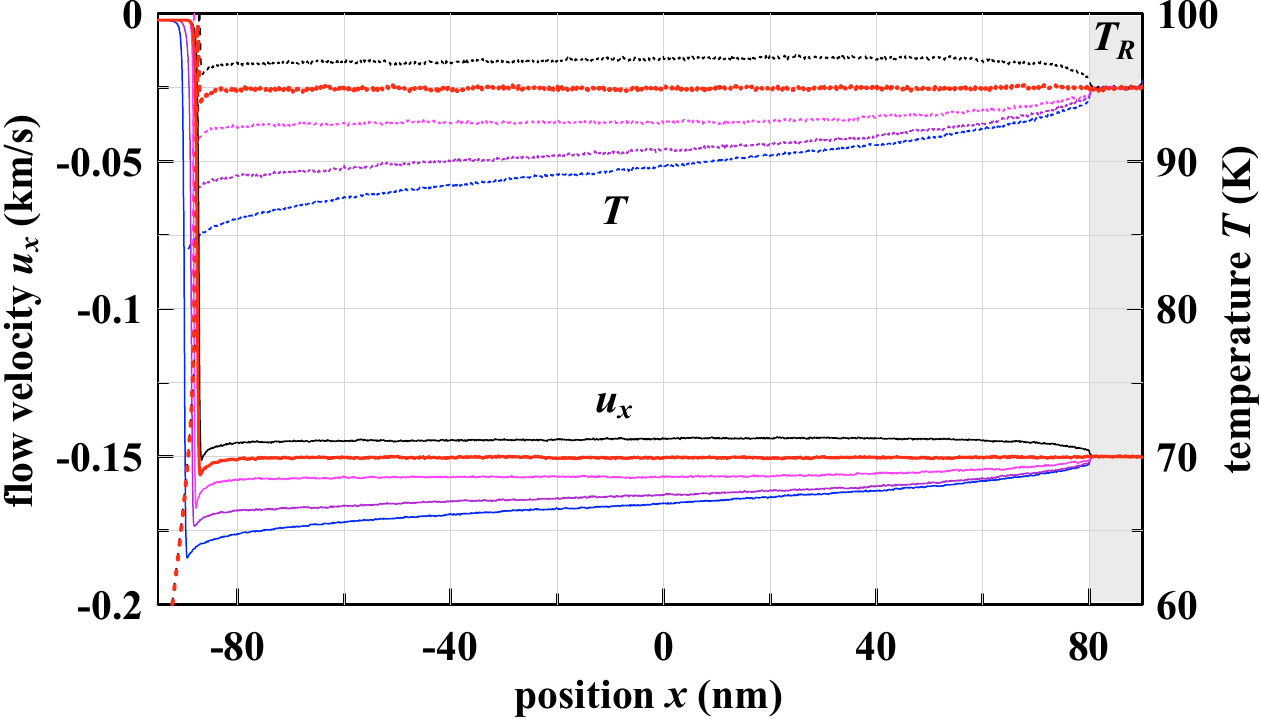}
\caption{\label{fig:Ux-T_wide} Temperature and flow velocity profiles obtained by MD simulation of subsonic condensation of gas flow with $u_{in}=-150\un{m/s}$ and $T_{in}= 95\un{K}$ applied within the right thermostat zone $T_R$. Red lines show profiles of a unique flow remained intact far from the condensation surface.}
\end{figure}

The pressure drop in the gas approaching the condensation surface intuitively seems to be a universal behavior, but calculations may also result in steady flow profiles with non-monotonic pressure behavior, even with a local excess of $P_x > P_\infty$, in which the gas can be decelerated and compressed. Such behavior is given in \cite{Bishaev:1973,Aoki:1989}

Condensation of a cold supersaturated vapor on a hot surface can also be realized, such unusual condensation at $T_s > T_\infty$ is demonstrated in Fig.~\ref{fig:Tx5_subson}. Here we show the longitudinal profiles of $T_x$ for steady condensation of gas with density $n_\infty=1.848\un{nm^{-3}}$, exceeding fivefold the density of saturated vapor. It appears that as the flow velocity decreases below $|u_\infty| < 126\un{m/s}$, condensation begins at the surface with $T_s > T_\infty=95\un{K}$, that is, hotter than the incoming gas. It should be noted that the perfect gas calculated by the momentum method condenses only on the surface and cannot homogeneously condense in a flow itself, unlike the real gas. Therefore, the considered case can be realized only if a  supersaturated gas source is sufficiently close to a condensation surface.

MD simulation of steady condensation of subsonic gas flow requires considerably more effort than simulation of supersonic condensation. This is due to the inevitable propagation of acoustic disturbances upstream, which complicates searching for a unique solution proving a steady profile of gas flow starting from infinity.

\begin{figure}[t]
\centering\includegraphics[width=1.\columnwidth]{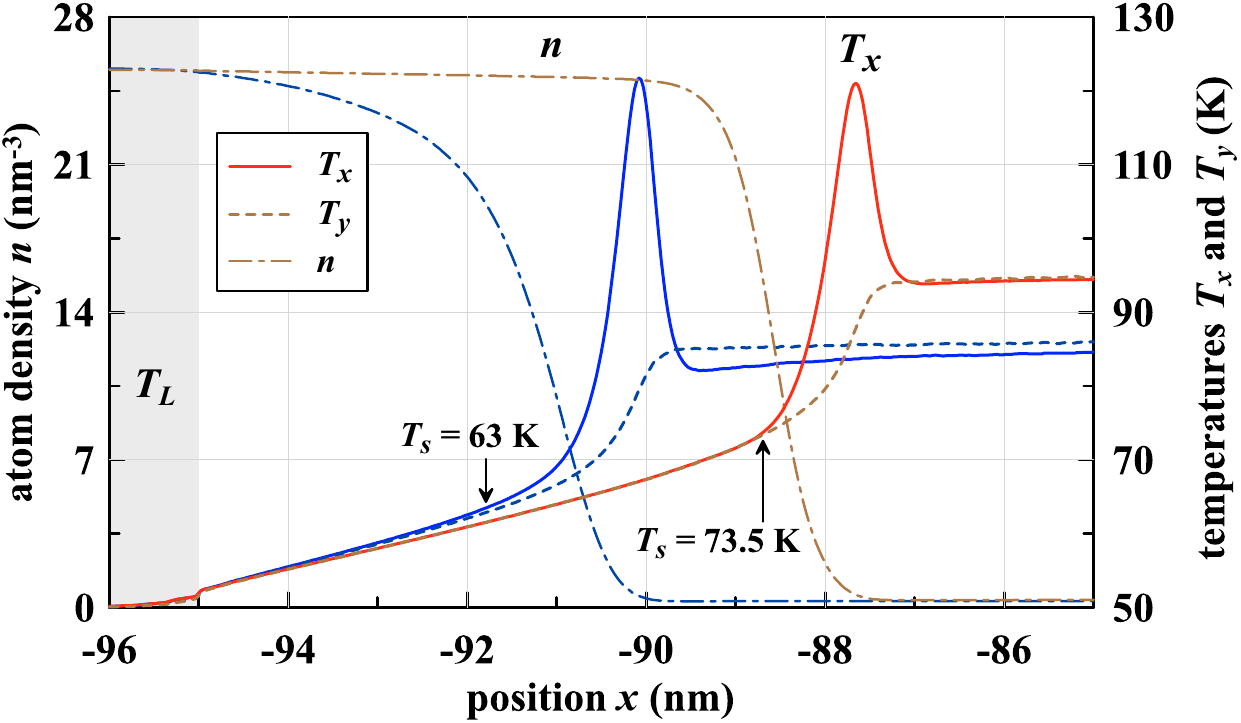}
\caption{\label{fig:Ux-T_cond} Temperatures and atom density profiles in the vicinity of condensation surface obtained by MD simulation of subsonic condensation. Red lines show the  steady profiles independent on distance from the condensation surface. Colors correspond to the same flows as on Fig.~\ref{fig:Ux-T_wide}.}
\end{figure}

To find such a solution, one has to perform several MD simulations, adjusting the temperature $T_s$ so that the flow inside the finite simulation domain is not perturbed far from the condensation surface. Figure~\ref{fig:Ux-T_wide} shows several calculated flow profiles, of which only the red profile can be considered a true solution, since it will remain unchanged as the right boundary is moved toward infinity.

The red-colored profile of this steady flow near the condensation surface is shown in Fig.~\ref{fig:Ux-T_cond}, where the blue profile of non-true solution depending on position of the right boundary is also given for comparison.
It is found from MD simulations that the unique surface temperature $T_s=73.5\un{K}$ is required to establish steady condensation of L-J gas flow with $u_\infty=-150\un{m/s}$, which agrees well with the $T_s=77.03\un{K}$ obtained by the momentum method for condensation of perfect gas of the same density and temperature.

\begin{figure}[t]
\centering\includegraphics[width=1.\columnwidth]{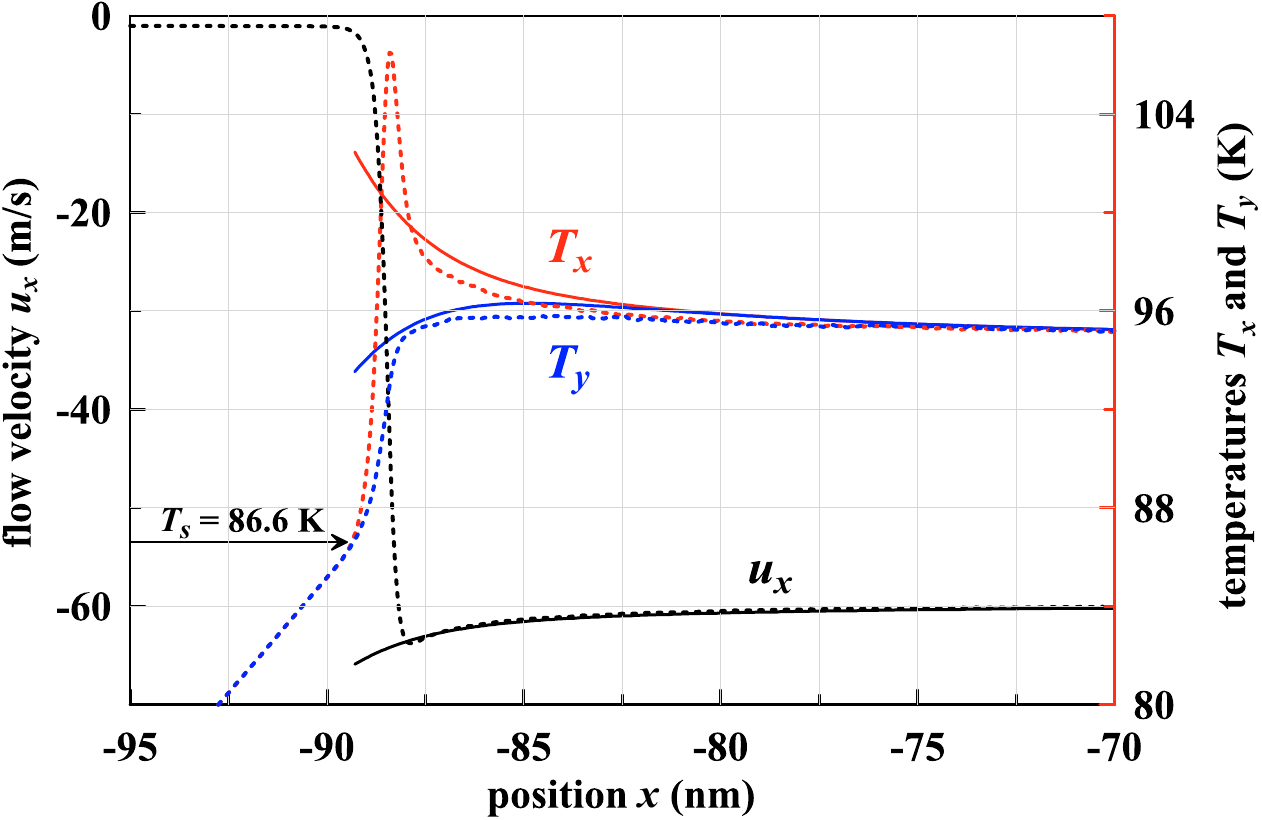}
\caption{\label{fig:Txy_MDvsBKE} Flow velocities, longitudinal and transverse vapor temperatures near condensation surface. Steady profiles are obtained by MD (dashed lines, $T_s=86.6\un{K}$) and BKE (solid lines, $T_s=87.99\un{K}$) for the same fixed subsonic flow with $u_\infty=-60\un{m/s}$ and $T_\infty=95\un{K}$.}
\end{figure}

A comparison of steady subsonic flow profiles found in MD simulations and calculated by the momentum method for the fixed flow velocity $u_\infty=-60\un{m/s}$ is shown in Fig.~\ref{fig:Txy_MDvsBKE}. Position of the condensation surface in the BKE profiles is aligned with the liquid phase boundary assigned to $T_s$ position determined from MD profiles.  As usual, the absence of a transient interface layer in the moment method makes it difficult to compare the calculated profiles directly with MD results, but the $T_s=87.99\un{K}$ obtained in BKE is close to $T_s=86.6\un{K}$ from MD. In contrast to the good agreement of the velocity profiles, noticeable differences are observed in the temperature profiles when approaching the interface layer.

\section{Elementary theory of steady condensation}
\label{sec:linear}

\begin{figure}[t]
\centering\includegraphics[width=1.\columnwidth]{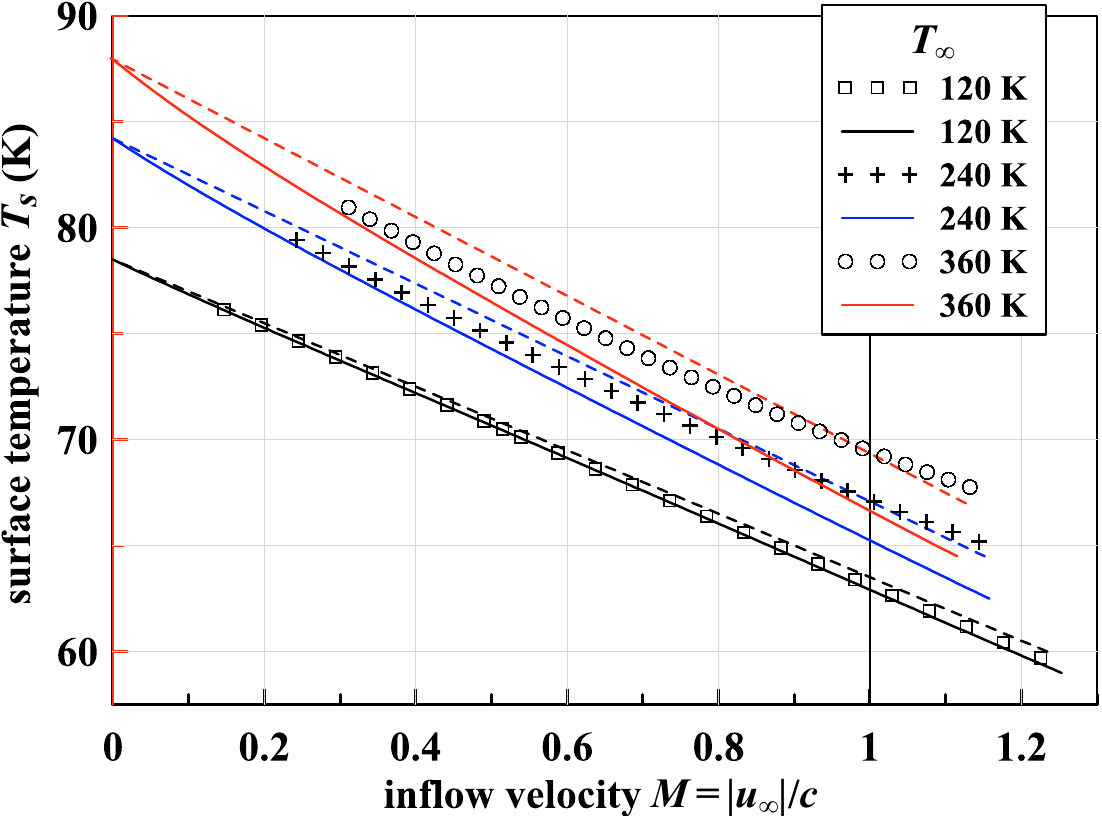}
\caption{\label{fig:Ts-U}
Liquid surface temperature as a function of the normalized gas flow velocity for three temperatures $T_\infty$ and the fixed concentration $n_\infty=0.048446\un{nm^{-3}}$. The data marked  by symbols are obtained from solutions of the moment equations, the solid curves are constructed using Eq.~(\ref{eq:LK_k1}), and the dashed lines correspond to the linearized Eq.~(\ref{eq:Uinflin}). The solutions for $M>1$ are not unique, but $T_s$ for a minimally perturbed flow in each case is shown.}
\end{figure}

\begin{figure}[t]
\centering\includegraphics[width=1.\columnwidth]{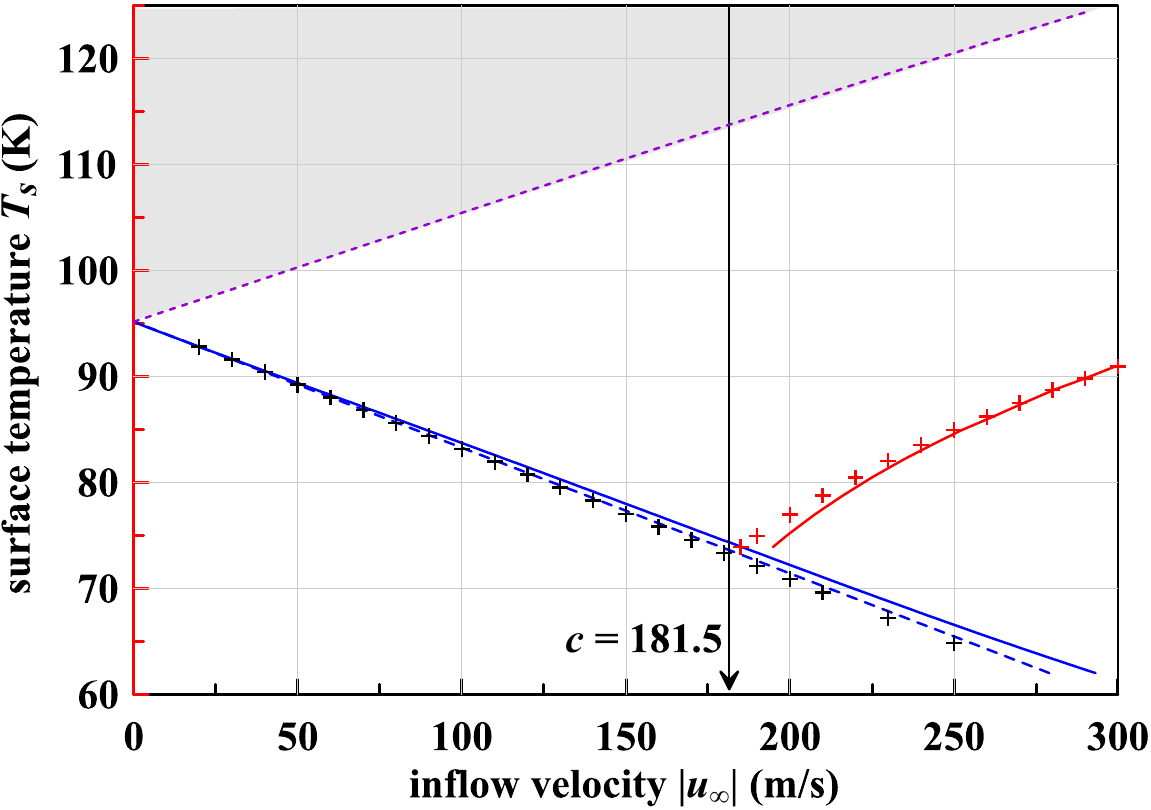}
\caption{\label{fig:Ts-U95}
Temperature of condensation surface as a function of the flow velocity of gas with the concentration $n_\infty=0.3696\un{nm^{-3}}$ of L-J vapor saturated at the temperature $T_\infty=95\un{K}$, and $c=181.5\un{m/s}$ is sound speed in this gas. Crosses point to data derived from solutions of the moment equations, the solid curves (blue and red) are based on Eq.~(\ref{eq:LK_k1}), and the blue dashed line corresponds to Eq.~(\ref{eq:Uinflin}). The red crosses show initial flow velocity obtained from the moment equations for a gas shocked by a standing SW. Condensation is not possible in the gray area bounded by the purple curve because of flow motion ceases by the strong outgoing SW. }
\end{figure}

The linearized theory of evaporation and condensation from/into semi-infinite space was constructed by D.~A.~Labuntsov and T.~M.~Muratova in 1969 \cite{Muratova:1969} in development of Labuntsov's work on evaporation-condensation \cite{Labuntsov:1967}. For relatively slow condensation a formula determining the condensation rate (mass flux density $j$) was obtained, which at condensation coefficient equal to one has the following form:
\begin{equation}
\label{eq:B1}
j=\rho_\infty u_\infty=\frac{5}{3} \frac{P_\infty-P_s}{\sqrt{2\pi R T_\infty}},
\end{equation}
where $T_\infty \approx T_s$ due to linearity of the problem.
To calculate the mass flux density for high-rate condensation, an improved formula has been proposed, which approximates the results of \cite{Labuntsov:1979} in the range of existence of one-dimensional steady flow profiles for subsonic velocities of inflow gas:
\begin{equation}
\label{eq:LK_k1}
  j=\frac{5}{3}\frac{P_\infty-P_s}{\sqrt{2 \pi RT_\infty}}\left[1+0.51\cdot \ln\left({\frac{P_\infty}{P_s}\sqrt{\frac{T_s}{T_\infty}}}\right)\right]
\end{equation}
The calculations using this formula agree quite well with the results obtained in \cite{Labuntsov:1979} by solving the Boltzmann kinetic equation by the momentum method, as demonstrated in Figs.~\ref{fig:Ts-U} and \ref{fig:Ts-U95}. Our calculations also show that the range of application of Eq.~(\ref{eq:LK_k1}) is much wider than expected, and it even includes supersonic flow condensation, as seen in Figs.~\ref{fig:Ts-U} as well as illustrated by the blue line in Fig.~\ref{fig:Ts-U95}. But it should be emphasized here that steady supersonic gas flows can condense on the surface at various $T_s$, i.e. there is no a unique solution of the moment equations at $M>1$. These figures show $T_s$ for a minimally perturbed supersonic flow, to which the boundary problem solutions converge as the right boundary is gradually moved toward infinity.

The application of the formula (\ref{eq:LK_k1}) is also possible to calculate the formation of a standing shock wave during condensation of a stationary supersonic flow, as shown in Fig.~\ref{fig:Ts-U95}. Here, to calculate $T_s$, the shock-compressed state of the perfect gas was first calculated for standing SW having speed equal to a modulus of the gas flow velocity. The resulting subsonic shock-compressed gas flow parameters were then used to calculate steady condensation using the momentum method (red crosses), and for comparison by Eq.~(\ref{eq:LK_k1}) -- see the red line in Fig.~\ref{fig:Ts-U95}.

At even higher $T_s$ the regime of complete cessation of condensation is realized, as discussed above in Section \ref{sec:supersonic}. The lower boundary of this regime is shown by the dashed purple line in Fig.~\ref{fig:Ts-U95}.

It can be demonstrated that the velocity $u_\infty$ of steady gas flow is a linear function of the condensation surface temperature. Let us write the gas flux density Eq.~\ref{eq:LK_k1} without a correction factor as:
\begin{equation}
j=\rho_\infty u_\infty= \frac{5}{3\sqrt{\pi}} \frac{P_\infty}
{\sqrt{2 RT_\infty}} \left(1-\frac{P_s}{P_\infty}\right)
\end{equation}
Then using perfect gas pressure $P=\rho RT$ we can write down the normalized flow velocity:
\begin{equation}
\frac{u_\infty}{\sqrt{2 RT_\infty}}= \frac{5}{6\sqrt{\pi}} \left(1-\frac{P_s}{P_\infty}\right)
\label{eq:Uinf}
\end{equation}
This formula relates the (usually known) gas parameters at infinity ($u_\infty$ and $T_\infty,P_\infty$) to the pressure $P_s$ of gas evaporating from the surface of condensed phase at temperature $T_s$, which determine a stationary solution of the condensation problem.

As was shown in Section \ref{sec:methods}, the concentration of saturated L-J vapor is represented by Eq.~
\ref{eq:Ns}, see Fig.~\ref{fig:Ns}. Then the pressure of corresponding perfect vapor is
\begin{equation}
P_s= k_B T \exp \left(a-\frac{T_v}{T}\right)
\end{equation}
Using this expression, we try to linearize Eq.~(\ref{eq:Uinf}). To do this, we match the gas pressure at infinity $P_\infty$ to the equivalent pressure of saturated vapor at some $T_e$ by the formula:
\begin{equation}
\label{eq:Pinf}
P_\infty= n_e k_B T_e = k_B T_e \exp \left(a-\frac{T_v}{T_e}\right).
\end{equation}
Then the pressure ratio $P_s/P_\infty$ can be written as:
\begin{equation}
\frac{P_s}{P_\infty} =
\frac{T_s}{T_e} \exp\left[\frac{T_v}{T_e} - \frac{T_v}{T_s}\right] =
\frac{T_s}{T_e} \exp\left[\left(1-\frac{T_e}{T_s}\right)\frac{T_v}{T_e}\right]
\end{equation}

Let's linearize this relation by leaving only the first linear term in the exponent expansion:
\begin{equation}
\frac{P_s}{P_\infty} \approx
\frac{T_s}{T_e} \left[1 + \left(1-\frac{T_e}{T_s}\right)\frac{T_v}{T_e}\right]
\end{equation}
After some simple transformations one can derive that
\begin{equation}
1-\frac{P_s}{P_\infty}\approx \left(1+\frac{T_v}{T_e}\right)\left(1-\frac{T_s}{T_e}\right)
\end{equation}

Thus, the linearized expression for the normalized flow velocity from Eq.~(\ref{eq:Uinf}) takes the form:
\begin{equation}
\label{eq:Uinflin}
\frac{u_\infty}{\sqrt{2 RT_\infty}} \approx \frac{5}{6\sqrt{\pi}} \left(1+\frac{T_v}{T_e}\right)\left(1-\frac{T_s}{T_e}\right)
\end{equation}
To use this formula, one must know the characteristic evaporation temperature $T_v=E_v/k_B$, and determine the equivalent temperature $T_e$ of saturated vapor having a pressure equal to the gas pressure at infinity $P_\infty$ by solving Eq.~(\ref{eq:Pinf}).

In general, if condensation is steady then Eq.~(\ref{eq:Uinflin}) can be interpreted as a relationship between four quantities: the condensation surface temperature $T_s$ and three flow parameters $u_\infty$, $T_\infty$, $T_e(P_\infty)$. Fixing any pair of these quantities, the other pair forms a simple dependence between them.

It should be emphasized that the presented deduction should by no means be considered rigorous, since the original Eq.~(\ref{eq:B1}) gives less accurate results than its linearized version Eq.~(\ref{eq:Uinflin}) when compared with ``accurate'' calculated values obtained from the stationary solutions of the moment equations. However, Figures~\ref{fig:Ts-U} and \ref{fig:Ts-U95} demonstrate that the heuristic formula (\ref{eq:Uinflin}) gives an unexpectedly good approximation to the results of the BKE calculations, which is almost as good as the more precise Eq.~(\ref{eq:LK_k1}) over a wide range of incoming flow velocities, including even supersonic flows. It should also be noted that Eq.~(\ref{eq:Uinflin}) loses its accuracy at approaching $T_s \to 0$ and the large Mach numbers.

\section{Conclusion}

By means of direct comparison of stationary condensations calculated by the moment method and with the help of molecular dynamic simulations, we have demonstrated good accuracy of the moment method at much less computational costs.
Surprisingly, we have also revealed that the approximate linear formulas quite accurately describe not only the weak condensation, but they are also applicable to condensation of supersonic flows.

Calculations of the surface temperature $T_s$ required for establishing stationary condensation of subsonic flow performed in our work by means of MD, BKE, or formula estimations allow us to distinguish the full condensation from partial condensation of the incoming flow. Full condensation occurs at surface temperature $T<T_s$, and partial condensation at $T>T_s$, when part of the flowing mass accumulates in front of the surface. For subsonic flow, this accumulation occurs above the blue line of stationary condensation in Fig.~\ref{fig:Ts-U95}, and for supersonic flow above the red line of stationary shock-compression condensation. In both modes, partial condensation ceases completely after generation of a sufficiently strong shockwave ahead of the condensation surface.

In practice, not only necessary conditions for condensation of gas flow are important, but also conditions necessary to prevent condensation. As we have shown, complete cessation of condensation is provided by a shock wave departing from the surface. For implementation of such a condition, the shockwave pressure must be high enough to stop the flow of shock-compressed gas to the condensation surface, which is realized if the pressure of evaporated gas is equal to the pressure of shock-compressed gas. In other words, the criterion for stopping condensation is determined by the shock Hugoniot of gas incoming to the surface.

\begin{acknowledgments}
This study was supported by the Russian Foundation for Basic Research, Grant No. 20-08-00342.
\end{acknowledgments}

\appendix*
\section{Equations of moment method}
\label{sec:appendix}

Let's define the reduced variables as
$\tau_1=\sqrt{T_1}$, $\tau_2=\sqrt{T_2}$, and $\upsilon_1=u_1/\sqrt{T_1}$, $\upsilon_2=u_2/\sqrt{T_2}$. The required functions are defined as follows
$\omega_1=\exp(-\upsilon_1^2)/\sqrt{\pi}$, $\omega_2=-\exp(-\upsilon_2^2)/\sqrt{\pi}$, and
$\Psi_1=1+\erf(\upsilon_1)$, $\Psi_2=1-\erf(\upsilon_2)$. Here the error function is
$\erf(z)=\frac{2}{\sqrt{\pi}}\int_0^z \exp(-x^2)dx$. Spatial derivatives of flow variables are written below as $n^{\prime}\equiv dn/dx$ and so forth.

The first moment equation constitutes the mass conservation (the continuity equation). It can be derived from the Boltzmann equation using $\varphi_1=1$, which gives the sum of left $(i=1)$ and right $(i=2)$ components as follows
\begin{equation}
\label{eq:a1}
\sum_i A_{1i}\bigl(\tau_i n_i^{\prime}+ n_i\tau^{\prime}_i\bigr)+ B_{1i} n_i \tau_i \upsilon^{\prime}_i = I_1
\end{equation}
where the 1st moment of the collision integral $I_1=0$, $A_{1i}=(\upsilon_i\Psi_i + \omega_i)/2$, $B_{1i} = \Psi_i$, and $i\in\{1,2\}$.

The second moment equation constitutes the momentum conservation law. It can be derived from the Boltzmann equation using $\varphi_2=\xi_x$, which gives
\begin{equation}
\label{eq:a2}
\sum_i A_{2i}\tau_i \bigl(\tau_i n_i^{\prime} + 2 n_i\tau_i^{\prime}\bigr) + B_{2i} n_i\tau_i^2 \upsilon_i^{\prime}= I_2,
\end{equation}
where the 2nd moment of the collision integral $I_2=0$, and $A_{2i} = \left(\upsilon_i^2 + 1/2\right)\Psi_i + \upsilon_i\omega_i$, $B_{2i} = 2(\upsilon_i\Psi_i + \omega_i)$.

The third moment equation constitutes the energy conservation law. It can be derived from the Boltzmann equation using $\varphi_3=\mathbf{\xi}^2$, which gives
\begin{equation}
\label{eq:a3}
\sum_i A_{3i}\tau_i^2\bigl(\tau_i n_i^{\prime}+ 3 n_i \tau_i^{\prime}\bigr)+ B_{3i} n_i\tau_i^3 \upsilon_i^{\prime}= I_3,
\end{equation}
where the 3d moment of the collision integral $I_3=0$, and
\begin{eqnarray}
\nonumber
A_{3i} &=& \left[(\upsilon_i^2+5/2)\upsilon_i\Psi_i+ (\upsilon_i^2+2)\omega_i\right]/2, \\
\nonumber
B_{3i} &=& \left[(3\upsilon_i^2+5/2)\Psi_i + 3\upsilon_i\omega_i\right]/2.
\end{eqnarray}

The forth moment equation is obtained for $\varphi_4=\xi_x^2$
\begin{equation}
\label{eq:a4}
\sum_i A_{4i}\tau_i^2\bigl(\tau_i n_i^{\prime} + 3 n_i\tau_i^{\prime}\bigr) + B_{4i} n_i\tau_i^3 \upsilon_i^{\prime} = I_4,
\end{equation}
where the 4th moment of the collision integral is
\begin{equation}
\nonumber
I_4 = \Bigl(\sum_i A_{1i} n_i\tau_i\Bigr)^2 - \sum_i A_{1i} n_i\tau_i^2 \sum_i n_i\Psi_i/2,
\end{equation}
and
\begin{eqnarray}
\nonumber
A_{4i} &=& \left(\upsilon_i^2+ 3/2\right)\upsilon_i\Psi_i + (\upsilon_i^2+1)\omega_i,\\
\nonumber
B_{4i} &=& 3\left(\upsilon_i^2+ 1/2\right)\Psi_i + 3\upsilon_i\omega_i.
\end{eqnarray}

The fifth momentum equation is obtained for $\varphi_5=\xi_x^3$
\begin{equation}
\label{eq:a5}
\sum_i A_{5i}\tau_i^3\bigl(\tau_i n_i^{\prime} + 4 n_i \tau_i^{\prime}\bigr) + B_{5i} n_i\tau_i^4 \upsilon_i^{\prime} =I_5,
\end{equation}
where the 5th moment of the collision integral is
\begin{eqnarray}
\nonumber
I_{5} = &\frac{3}{2}& \sum_i A_{1i} n_i\tau_i \sum_i n_i\tau_i^2 \left[(\upsilon_i^2+1)\Psi_i+ \upsilon_i\omega_i\right] - \\
\nonumber
&\frac{3}{4}& \sum_i n_i\Psi_i \sum_i n_i\tau_i^3 \left[(\upsilon_i^2 + 1)\upsilon_i\Psi_i + (\upsilon_i^2+ 1/2)\omega_i\right],
\end{eqnarray}
and
\begin{eqnarray}
\nonumber
A_{5i} &=& \left(2\upsilon_i^4+6 \upsilon_i^2 + 3/2\right)\Psi_i+ \left(2\upsilon_i^2 + 5\right)\upsilon_i\omega_i, \\
\nonumber
B_{5i} &=& 4\left(2\upsilon_i^2 + 3\right)\upsilon_i\Psi_i + 8(\upsilon_i^2+1)\omega_i.
\end{eqnarray}

The sixth momentum equation is obtained for $\varphi_6=\xi_x \mathbf{\xi}^2$
\begin{equation}
\label{eq:a6}
\sum_i A_{6i}\tau_i^3 (\tau_i n_i^{\prime} + 4 n_i \tau_i^{\prime}) + B_{6i} n_i\tau_i^4  \upsilon_i^{\prime} =I_6,
\end{equation}
where the 6th moment of the collision integral is
\begin{eqnarray}
\nonumber
I_6 &=& \sum_i A_{1i}n_i\tau_i \sum_i n_i \tau_i^2 \left[\left(\upsilon_i^2+ 5/2\right)\Psi_i+ \upsilon_i\omega_i\right] - \\
\nonumber
& &\frac{1}{2} \sum_i n_i\Psi_i \sum_i n_i \tau_i^3 \left[\left(\upsilon_i^2 + 5/2\right)\upsilon_i\Psi_i+ (\upsilon_i^2+2)\omega_i\right]
\end{eqnarray}
and
\begin{eqnarray}
\nonumber
A_{6i} &=& \left(2 \upsilon_i^4+8 \upsilon_i^2+ 5/2\right)\Psi_i + \left(2 \upsilon_i^2 + 7\right)\upsilon_i\omega_i, \\
\nonumber
B_{6i} &=& 8\left(\upsilon_i^2 + 2\right)\upsilon_i\Psi_i + 4\left(2\upsilon_i^2 + 3\right)\omega_i.
\end{eqnarray}

\bibliography{condensation}

\end{document}